\PassOptionsToPackage{table}{xcolor}
\documentclass[sigconf,nonacm]{acmart}

\usepackage{booktabs}
\usepackage{graphicx}
\usepackage{multirow}
\usepackage{cancel}
\usepackage{amsmath}
\usepackage{tabularx}
\usepackage{xcolor}
\usepackage{soul}
\usepackage{enumitem}
\usepackage{enumitem}
\usepackage{bm}

\setlist[description]{leftmargin=0cm,labelindent=0cm}

\AtBeginDocument{%
  \providecommand\BibTeX{{%
    \normalfont B\kern-0.5em{\scshape i\kern-0.25em b}\kern-0.8em\TeX}}}

\copyrightyear{2025}
\acmYear{2025}
\setcopyright{cc}
\setcctype{by-nc-sa}
\acmConference{Access InContext Workshop @ CHI’25}{April 26}{Yokohama, Japan}
\acmBooktitle{Access InContext Workshop @ CHI’25, April 26, 2025, Yokohama, Japan}

\begin{document}


\title{EqualMotion: Accessible Motion Capture for the Creative Industries}

\author{Clarice Hilton}
\email{c.hilton@gold.ac.uk}
\orcid{}
\affiliation{
  \institution{Goldsmiths, University of London}
  \city{London}
  \country{UK}
}

\author{Kat Hawkins}
\affiliation{
  \institution{Coventry University}
  \city{Coventry}
  \country{UK}
}

\author{Phill Tew}
\email{me@philltew.com}
\author{Freddie Collins}
\email{freddiecollins@x-io.co.uk}
\author{Seb Madgwick}
\email{sebmadgwick@x-io.co.uk}
\affiliation{
  \institution{x-io Technologies}
  \city{Bristol}
  \country{UK}
}

\author{Dominic Potts}
\email{dominic.potts@uwe.ac.uk}
\orcid{0000-0002-4607-289X}
\author{Tom Mitchell}
\email{tom.mitchell@uwe.ac.uk}
\orcid{0000-0003-0944-4989}
\affiliation{
  \institution{University of the West of England}
  \city{Bristol}
  \country{UK}
}

\renewcommand{\shortauthors}{Hilton et al.}


\begin{abstract}
Motion capture technologies are increasingly used in creative and performance contexts but often exclude disabled practitioners due to normative assumptions in body modeling, calibration, and avatar representation. EqualMotion introduces a body-agnostic, wearable motion capture system designed through a disability-centred co-design approach. By enabling personalised calibration, integrating mobility aids, and adopting an inclusive visual language, EqualMotion supports diverse body types and movement styles. The system is developed collaboratively with disabled researchers and creatives, aiming to foster equitable participation in digital performance and prototyping. This paper outlines the system's design principles and highlights ongoing case studies in dance and music to evaluate accessibility in real-world creative workflows.
\end{abstract}

%% The code below is generated by the tool at http://dl.acm.org/ccs.cf.


\begin{teaserfigure}
   \vspace{-3mm}
   \centering
   \includegraphics[width=\textwidth]{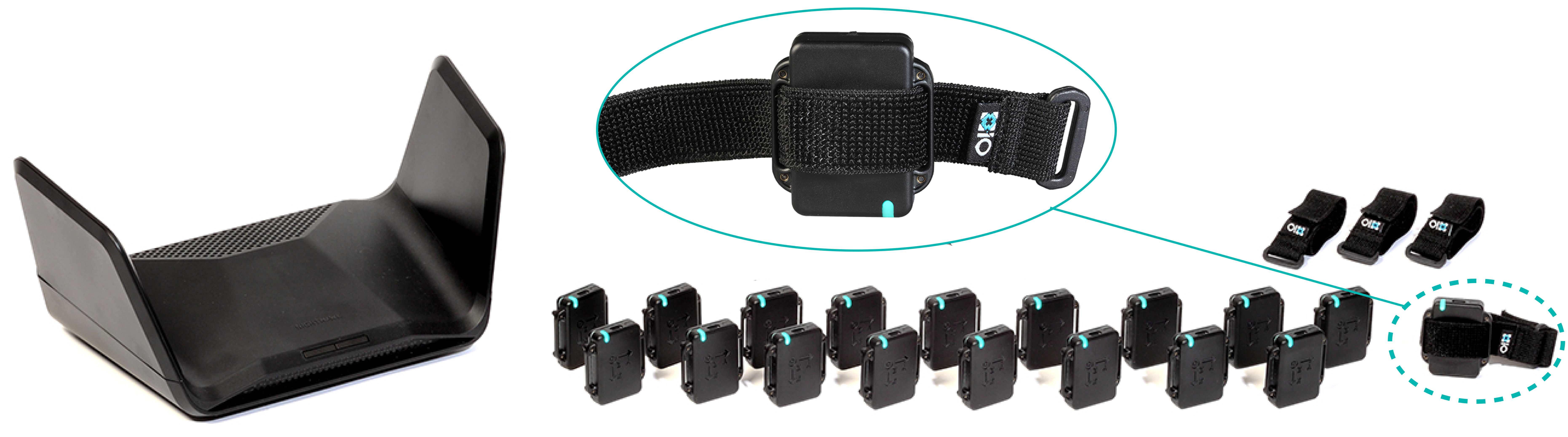}
   \vspace{-3mm}
    \caption{The EqualMotion motion capture system. The suit is constituted by an array of networked x-IMU3s that can be mounted on different parts of the user's body and on mobility aids such as wheelchairs and prosthetics to capture user movement in 3D.}
    \Description{A black plastic flight case that is open showing the contents. There is black rigid foam inside, with various cut-outs with different pieces of hardware inside. In the top left of the case are two powerbanks with numerous USB ports. To the right of this are 18 x-IMU3 modules sitting vertically and in rows in the case. Below this is a black network router used to connect the IMUs together. One of the x-IMU3s is annotated with a line and bubble, showing a close-up of the module. The x-IMU3 module is a black and turquoise microcontroller in a black plastic housing with a turquoise LED and usb-c socket.}
\label{fig:teaser}
\end{teaserfigure}

\maketitle

\section{Motivation}
%% Clarice: I'm not sure about this framing as the most important part of mocap. 
%% Tom: Remove this paragrph?
%Accurately measuring and recreating the motion of the human body is important for achieving authentic and life-like animation of avatars and the physical interaction of participants within virtual environments.

% \begin{figure}
%     \centering
%     \includegraphics[width=1\linewidth]{source/Figures/EM_Kit-03-03.png}
%     \caption{The EqualMotion motion capture system. The suit is constituted by an array of networked x-IMU3s that can be mounted on different parts of the user's body and on mobility aids such as wheelchairs and prosthetics to capture user movement in 3D.}
%     \Description{A black plastic flight case that is open showing the contents. There is black rigid foam inside, with various cut-outs with different pieces of hardware inside. In the top left of the case are two powerbanks with numerous USB ports. To the right of this are 18 x-IMU3 modules sitting vertically and in rows in the case. Below this is a black network router used to connect the IMUs together. One of the x-IMU3s is annotated with a line and bubble, showing a close-up of the module. The x-IMU3 module is a black and turquoise microcontroller in a black plastic housing with a turquoise LED and usb-c socket.}
%     \label{fig:EM_Kit}
% \end{figure}

Motion capture (or mocap) \cite{menolotto_motion_2020} is the process of digitally recording body motion and is an important tool for animators \cite{sharma_use_2019}, roboticists \cite{field_motion_2009}, and sports scientists \cite{suo_motion_2024}. It is increasingly used in the context of creative, mixed-reality performances that connect both physical and virtual worlds \cite{kitagawa_mocap_2020}.

However, affordable commercial mocap systems are inaccessible at a foundational level for disabled creative practitioners and researchers~\cite{Harvey_2024_TheCadaver}. Due to the normative assumptions about the bodies of the users they are intended to capture, prototyping with such systems for disabled users is either not possible or requires system hacking. Furthermore, mocap systems are often trained on representations of normative bodies and often require users to wear specific clothing that is designed for normative bodies~\cite{Vicon_2025, Qualisys_2025, Rokoko_2025, Xsens_2025}. Consequently, these systems exclude disabled people from fully participating in using and prototyping with motion capture.

There is now growing interest in the design of accessible motion capture systems, see for example WheelPoser, a wearable system designed to estimate the body pose of wheelchair users \cite{li_wheelposer_2024}.

We introduce EqualMotion\footnote{\url{https://www.myworld-creates.com/funding/funded-projects/equalmotion-accessible-motion-capture-for-the-creative-industries/}} (shown in ~\autoref{fig:teaser}), a project that takes a disability-centred co-design approach to design a wearable, IMU-based mocap system in collaboration with disabled researchers and creative practitioners. From our work so far, the main points for inaccessibility in existing mocap systems are:

\begin{enumerate}
    \item Calibration procedures require inaccessible poses or movements (for example the `T pose').
    \item Tracking assumes a normative body.
    \item Participants are represented visually using a fixed 3D avatar that represents a normative body.
    \item Mobility aids, such as wheelchairs are often not considered or captured. 
\end{enumerate}

These limitations result in systems that prevent disabled users from fully participating in the creation of virtual content entirely, or creating awkward or potentially harmful working scenarios; for example, in this video a wheelchair tennis athlete is represented by a standing avatar~\footnote{NTR (Netherlands) ``\textit{The ideal wheelchair}'', see 8:50: ~\url{https://ntr.nl/Focus/287/detail/De-ideale-rolstoel/VPWON_1344843}}.

\section{EqualMotion Design}
The project is guided by the social model of disability, adopting a `nothing about us without us' approach \cite{charlton_nothing_1998}. 
The EqualMotion research team includes disabled and non-disabled researchers and the project engages with disabled creative practitioners in the co-design of a more accessible motion-capture system. Through a disability-centred methodology and workshopping with expert practitioners, we have identified the following design features of EqualMotion, which we aim to explore further:

\paragraph{Body Agnostic Motion Capture}
Traditional motion capture systems make assumptions about the topology of the body and how different body parts relate to one another. These assumptions in the sensor calibration and virtual representation are foundational, making the system fundamentally inaccessible.

The EqualMotion system takes a modular, body-agnostic approach, building a user’s motion profile from a central `root' IMU module. Additional nodes are then added, labelled using self-identified language, and then mapped based on the user’s unique range of motion. This enables accurate representation of diverse bodies, including mobility aids. By centring personalisation, EqualMotion allows motion capture to reflect an individual’s natural movement—for example, accommodating a user whose expressive motion is concentrated in a single arm.

% - problem - 
%Common motion capture systems have particular assumptions about the user's body and its range of motion, such as having two arms and two legs. Systems rely on these assumptions for calibration of sensors across a user's body. However, these built-in assumptions about a user's body are a foundational problem leading to lacking consideration of non-normative bodies in both calibration and virtual representation.

% - solution space
%The EqualMotion system takes a modular, body-agnostic approach in which a user's body is built outwards from a `root' IMU module, and subsequent nodes are added, labelled, and related to other nodes defined by the user's range of motion. This not only enables the creation of non-normative representations of bodies but also incorporates extensions to the body in motion capture such as mobility aids. The body-agnostic principles of EqualMotion also allow for motion that is highly personalised to a user's natural range of motion, for example, one user's expressive movements may be entirely located in one arm and so the motion capture system can be constructed around this. 

\paragraph{Personalised Calibration}
Traditional motion capture systems require users to adopt predefined calibration poses, such as the T-pose or A-pose~\cite{ROBERTLACHAINE201780}, to orient IMU nodes and generate a skeleton model. These poses are expected to be \textit{comfortable}, \textit{neutral}, and \textit{repeatable}, yet they may be difficult or impossible for disabled users to maintain. EqualMotion is designing personalised calibration, allowing users to choose and create their own poses, such as seated or supine positions.

Additionally, a dataset of accessible calibration poses and profiles will be created for general use of EqualMotion. While these profiles cover a finite number of access requirements, such as wheelchair users, they will create a more accessible onboarding for rapid prototyping with the option of creating a bespoke profile later.

%This is suitable for iterative prototyping or only approximate tracking, while still catering for a wider set of non-normative bodies than conventional mocap systems

% - problem - poses are inaccessible, labels make assumptions.
%Current mocap systems rely on users adopting a known pose during calibration, orientating the individual IMU nodes to create a skeleton, usually the `T-pose' or `A-pose' ~\cite{}. From the user's perspective, these poses must be \textit{comfortable}, \textit{neutral}, and easily \textit{repeatable}. For disabled users, these poses may be uncomfortable to adopt for any length of time or in some cases impossible. Furthermore, the labelling of a user's body in mocap software often makes assumptions about the user's body and might not always account for non-normative bodies.

% - solution space
%EqualMotion will support diverse and personalised calibration profiles allowing users to adopt calibration poses that suit their access requirements, such as seated and prone poses, as well as label their virtual representation with language they deem appropriate.

\paragraph{Mobility Aid Integration}
Commercial mocap systems rarely account for mobility aids such as wheelchairs, crutches, or prosthetics, despite their integral role in users’ movement and expressivity~\cite{zhang2025inclusive}. While some aids, like prosthetic limbs, can be represented in existing systems, others (e.g. wheels) have unique kinematics that are often overlooked.

Standard IMU mounting methods may not be practical for certain mobility aids, and the materials used in these devices can interfere with tracking. For example, ferrous metals can disrupt IMU magnetometers, which play a key role in pose estimation.

EqualMotion will integrate mobility aids into its core tracking principles, incorporating specialized tracking for wheels, mitigating material-based interference, and offering diverse mounting solutions to accommodate a wide range of mobility aids.

% - problem - unique kinematics, mounting, and tracking inteference
%Commercial mocap systems do not consider capturing the motion of mobility aids such as wheelchairs, crutches, or prosthetics that are often considered a part of the user's body and contribute to their expressivity and range of motion~\cite{}. While some mobility aids can more easily be represented in commercial mocap systems, for example, limb prosethics, some mobility aids have unique kinematics, such as wheels, whose movement is not often considered. 

%Conventional mounting methods for IMU modules may not be practical for certain mobility aids and the materials commonly used in mobility aids may have implications on the tracking quality of mocap suits. For example, close contact with magnetic metals causes interference with IMU's magnetometers which often informs the pose of a body. 

% - solution space
%EqualMotion considers unique aspects of mobility aids in its foundational tracking principles, for example, tracking primitives for wheels accounting for the unique kinematics, accounts for interference caused by mobility aid materials, and provides a diverse set of mounting methods to cover a wide range of mobility aids as possible. 

\paragraph{From Abstract to Specific Body Representations}
Normative virtual body representations in mocap systems often force users into avatars that do not reflect their own bodies. As ~\citeauthor{Gerling_2021_Avatars} describe representations assume ``a ‘corporeal standard’ (i.e., an ‘ideal’, non-disabled human body), and fails to adequately accommodate disabled people''~\cite{Gerling_2021_Avatars}. Additionally, there is a demonstrable preference for disabled users to embody more accurate representations of their bodies~\cite{zhang_diary_2023, mack_towards_2023}. 

EqualMotion is designing a non-assumptive visual language that gradually transitions from ambiguous abstraction to disambiguate specificity during calibration, easing users into embodiment. For example, before calibration, the user's body nodes are visually indistinguishable within a constellation of other nodes. As calibration progresses and initial movements are detected, unassociated nodes fade away while relevant ones become more prominent, ultimately forming an abstract yet personalized line-and-node representation of the user’s body.

\section{Fostering Inclusive Co-design}
EqualMotion looks to address access limitations in existing mocap systems by centring disabled researchers and practitioners in the development process of open-source software tools. The previously discussed design principles were synthesised through workshops with critical-disability researchers, HCI researchers, disabled movement practitioners, and motion capture engineers. While these offer a promising direction of development for EqualMotion, we also endeavour to assess the developed technology in the contexts where it can make the most significant impact. 

The development programme includes case study activities, that anticipate, highlight and address identified technical and access issues, informing the development of EqualMotion in context. These consist of two case studies, each focussing on a different movement-based practice: dance and music.

In the case studies, we aim to integrate EqualMotion into practitioners' existing workflows to allow full participation in the co-design and evaluation of EqualMotion. For example, in the music case study, our lead practitioner uses particular software tools, and to support this, we will develop software bridges between these different applications. 

As we further develop EqualMotion through these case studies, there are still open questions about both the co-design methodology of the case studies, and also the accessibility of the EqualMotion system itself:

\begin{itemize}
    \item How do we disrupt the usual subjectivities of disabled/non-disabled subject/researcher/self/other? What do we uncover about the networks of power in doing so?
    \item What can we learn from a prolonged period of a disabled performer using the suit within their practice?
\end{itemize}

\begin{acks}
The authors acknowledge support from UKRI (grant Number MR/X036103/1), MyWorld, and InnovateUK. 

We would like to dedicate this paper to Kat Hawkins, who devastatingly passed away during this project. Their brilliant methodologies, critiques, design and input were central in the development of the ideas and research described in this paper.
\end{acks}

\balance
\bibliographystyle{ACM-Reference-Format}
\bibliography{bibliography}

\end{document}